\documentclass[aps,twocolumn,showpacs,epsfig,epsf,amssymb,showkeys]{revtex4}
\usepackage{graphicx}
\usepackage{subfigure}
\begin{document}

\title{Self assembly of monodisperse CdS nano-cylinders with a pore.
}

\author{J. Kiruthiga$^1$, Apratim Chatterji$^2$}
\email{apratim@iiserpune.ac.in}
\affiliation{
$^1$ Department of Physics, Pondicherry University, Puducherry-605014,India.  \\
$^2$IISER-Pune, 900 NCL Innovation Park, Dr. Homi Bhaba Road,  Pune-411008, India.
}
\date{\today}
\begin{abstract}

We present our conclusions of the investigation of 
the self assembly and growth of an {\em array} of  $CdS$
nanotubes: a consequence of a fine balance of directed motion, diffusion and aggregation of reacting 
${\rm Cd^{+2}}$ and ${\rm S^{-2}}$ ions.
In a previous communication \cite{kiruthiga}, we identified the mechanism of a unexpected  growth of a 
very uniform $CdS$  nano-cylinder from the end of a nano-channel. Furthermore, 
the cylinder had a pore along the axis but were closed at one end. This unique phenomenon of
self assembly of  {\em monodisperse} CdS nano-cylinders had been  observed in a rather simple experiment 
where two chambers containing 0.1 M ${\rm Cd Cl_2 }$ and 0.1 M ${\rm Na_2 S}$ solutions were joined by an array 
of anodized aluminium oxide (AAO) nano-channels \cite{shouvik}.    
Interestingly, the growth of CdS  nano-tubes was observed only in the  ${\rm Na_2 S } $-chamber. 
Our previous study  focussed on identifying the principles governing the growth 
of a single nano-tube at the exit point of a single AAO-nano-channel.
In this communication,  we identify factors affecting the self-assembly of a nano-tube in  
the presence of neighbouring nano-tubes growing out an array of closely spaced AAO nano-channel exits, 
a study closer to experimental reality.  Our model is not $Cd^{+2}$ or $S^{2-}$ specific, and our 
conclusions suggest that the experimental scheme can be extended to self assemble a general class of 
reacting-diffusing  A and B ions with  A (in this case ${\rm Cd^{+2}}$) selectively migrating 
out from a nano-channel.  In particular,  we note that after the initial prolonged  growth of nanotubes, 
there can arise  a severe deficiency of  B-ions (${\rm S^{-2}}$) ions near the AAO-nano-channel exits, the points 
where the reaction and aggregation occurs to form the $CdS$ nanotube, 
thus impeding further growth of uniform CdS nano-tubes.
We further identify the parameters which can be tuned to obtain an improved crop of monodisperse nanotubes.
Thereby  we predict the necessary characteristics of reacting systems which can be 
self assembled using suitable adaptations of experiments used to grow CdS cylinders.


\end{abstract}

\maketitle

Controlled self assembly of micron to nanometer sized structures of different morphologies 
has been at the forefront of research interests for over a decade spanning disciplines of 
physics, chemistry and even biology \cite{israel,vermant,witten,biop,einax,opto}.  Recently, a very simple experiment produced the unexpected 
growth of very uniform Cadmium sulphide (CdS) nano-structures of rather unique morphology  as
was reported by Varghese and Datta \cite{shouvik}. The authors took $0.1$ M ${\rm CdCl_2}$ 
and $0.1$ M ${\rm Na_2S}$ solutions in two different chambers and allowed them to come into contact 
with each other through some Anodized Aluminium oxide (AAO) nano-channels. The diameter of 
AAO nano-cylinders was was varied from $20$ nm to $100$ nm. The radial dimension of the AAO nano-channels
are significantly larger than the ionic dimensions as well as the Bjerrum length ($ \sim 7 \AA$ at room temperature),
thereby the authors expected the nano-channels to get clogged by CdS precipitate formed 
by diffusing and reacting ${\rm Cd^{+2}}$ and ${\rm S^{-2}}$ ions  inside the AAO nano-channel.

Contrary to their expectations, cylindrical CdS nano-tubes with a pore along the center of the cylinder 
but closed at one end were found to grow outwards from the ends of the AAO nano-channels. The diameter 
of the CdS nano-cylinders were of the same order as that of the AAO nano-channels ({\bf N-C} for brevity), SEM photos of the 
CdS nano-cylinders can be found in reference \cite{shouvik}.  Furthermore, the CdS nano-structures 
were found to grow in the ${\rm Na_2 S}$ chamber only and never in the ${\rm CdCl_2}$ chamber.
The ${\rm CdS}$ nano-tubes ({\bf N-T} in short) with a pore continued to grow  in the ${\rm CdCl_2}$ chamber even if the 
surface charge on the  AAO N-C was reversed during the preparation, clearly establishing that 
the  selective migration of ${\rm Cd^{+2}}$ from the ${\rm CdCl_2}$ to ${\rm Na_2S}$ chamber is not a 
simple electrostatic potential effect in the presence of ionic screening or otherwise. Traces of 
${\rm CdS}$ are not found in the ${\rm CdCl_2}$  chamber indicating ${\rm S^{-2}}$ ions do not migrate to 
the ${\rm CdCl_2}$ chamber through the AAO N-C. However as a control experiment, if the ${\rm CdCl_2}$ solution is 
replaced by a chamber of pure water connected to the $0.1$ M  ${\rm Na_2 S}$ solution, one does
detect  ${\rm S^{-2}}$ ions in chamber containing pure water. One can conclude that it is not just AAO-specific properties 
which  prevent  ${\rm S^{-2}}$ ions from migrating to the ${\rm CdCl_2 S}$ chamber. Further experiments 
by concerned researchers are needed before one can chose one of the following scenarios (amongst many others) 
as a possible cause of selective transport of  ${\rm Cd^{+2}}$: (a) capillary action induced by chemical potential 
difference in the two chambers leading to directed motion of ${\rm CdCl_2}$ solution 
(b) formation of different-sized large hydrated ${\rm S^{-2}}$ and 
${\rm Cd^{+2}}$ ion-clusters  with very different diffusivities (c) AAO N-C induced low 
density of ions inside the channel, thereby, increasing the effective charge screening length.

The other surprising aspect of the experimental observations is the unexpected morphology 
of the $CdS$ nanotubes formed. What could be the physical conditions at the exit of the AAO-nanochannel
that a cylinder with a pore along the center and closed at one end would be formed, each cylinder jutting out 
from the AAO N-C ends ? This question was investigated in a previous paper of ours, where the main focus 
was on the growth of a single N-T in the ${\rm Na_2 S}$ chamber by exiting ${\rm Cd^{+2}}$ 
from a isolated AAO N-C. These ideas are applicable to a general class of reacting-diffusing and aggregating-precipitating
chemical species, and hence for the rest of the discussion we shall refer of ${\rm Cd^{+2}}$ ions are A,
${\rm S^{-2}}$ ions as B; thereby $ A + B \rightarrow C$, where CdS $\equiv$ C. $Na^{+}$ ions remains passive
in the model of the process and will not be explicitly considered in the rest of the paper.

The key ingredient to understand the observed morphology of cylinders with a pore is the assumption that 
A particles exit the  N-C with a finite velocity to enter  a bath of 
B-ions. Such selective transport of fluids through a nano-pore is not a unreasonable assumption to make \cite{majumdar}. 
These A-ions then meet the diffusing B-ions to form C, which in turn diffuse around a bit before
they find an appropriate site  in the forming C-aggregate. At times $t=0 +$ of the growth process, 
A ions meet lots of B, thereby react  and aggregate to form a plug of C which is pushed ahead by the pressure of fluid,
exiting the N-C. The region between the plug and the N-C exit becomes B-ion scarce thereafter, and 
is filled in rapidly by the exiting fluid containing A-ions. These then diffuse out to meet the diffusing-in B ions. $A + B \rightarrow C$ 
occurs and aggregation of C forms the walls of the N-T cylinder with a closed plug at one end.
These ideas were implemented in a lattice model by us and we obtained cylinders with a pore 
exactly as seen in experiments. A detailed description can be obtained from reference \cite{kiruthiga}.

The main focus of this rapid communication is the growth of multiple C-nanotubes (N-T) from an array of exit points
of AAO N-C. The presence of neigbouring growing N-Ts adds an extra complication to into the phenomenon.
The supply of reacting B-ions gets severly depleted near the AAO N-C exit as the reaction proceeds.
Diffusion is the principal pathway by which B-ions can move from bulk into the regions near the 
reaction points close to the N-C exits. In this paper we explore the consequences of the slow diffusion 
of B and other  conditions and criteria which effectively gets imposed  to have well formed N-Ts  of C particles,
with  few C-deficient regions within cylinder-walls. This in turn will help experiments to identify
and fine tune conditions to self assemble structures from different chemical species separated by a suitably chosen nano-channel. 

We very briefly describe the model of self-assembling N-T, this will also help define the various physical quantities
relevant for our model. The ${\rm Na_2 S}$ chamber is modelled as a 3-D lattice of size $L_x \times  L_y \times  L_z$,
and A and B can reside on only discrete points of a lattice. A and B are self-avoiding, however, a diffusing A (or B)
ion can hop into a site occupied by B (or A) which then turns into C particle. The probability that a A,B or C 
particle will attempt a random hop to a neighbouring vacant site (mimicking diffusion) is given 
by $D_A, D_B$ and $D_C$ respectively. The exit points of the N-C (explicitly not modelled) are at $x=0$, A enters 
the lattice from different areas of dimension $2 r_{NC}   \times 2 r_{NC}$, in turn each of these square exit-areas 
form a square lattice. This corresponds to N-Cs being arranged in a square lattice 
in our model for ease of computation, though one can in principle also work with a triangular lattice. 
The center of from the center of a square to the center of the nearest neighbour
square is $d_c$, and the distance between neighboring N-C walls is $S_{cyl} = d_c - 2 r_{NC}$. $L_y \times L_z$
are chosen such that PBC can be maintained in $y$, $z$ direction as $S_{cyl}$ is varied. 

Flow of A is modelled by making A hop a lattice constant $a$ to the 
right (i.e.$+\hat{x}$ direction) every iteration (time $\tau$) if a vacant site or a site occupied by B is available.
Thus the distance $a$ hopped in one time step $\tau$ sets the length and time scale of the model. If the lattice 
site on the right of a about-to-hop A is occupied by another A ion, then the first A tried to ``flow'' around 
the A on the right by taking a random step in $\pm \hat{y}$ or $\pm \hat{z} $ direction if suitably vacant. At each time-step,
A-ions are replenished with probability $p_A$ at each of lattice-points which constitute the square exits of the  N-C at $x=0$. 
The initial density of B ions in the lattice is set to be $\rho_B$, i.e., $\rho_B$ is fraction of all lattice points 
are occupied by B at time $t=0$. Note however, that if the exit of N-C are already occupied by A, then no new A-s effectively
get introduced at the lattice sites at $x=0$, till they become unoccupied by a hop of A-s in the $+\hat{x}$ direction.
The number of interations $N_{it}$ are chosen such that the average length of the N-T at the end of simulation is $\approx 200 a$, 
i.e. $p_P \times N_{it} =200$; thus lower values of $p_P$ correspond to longer simulation runs. Correspondingly $L_x =1000a$ 
for all our runs.

\begin{figure}
\includegraphics[width=0.45\columnwidth]{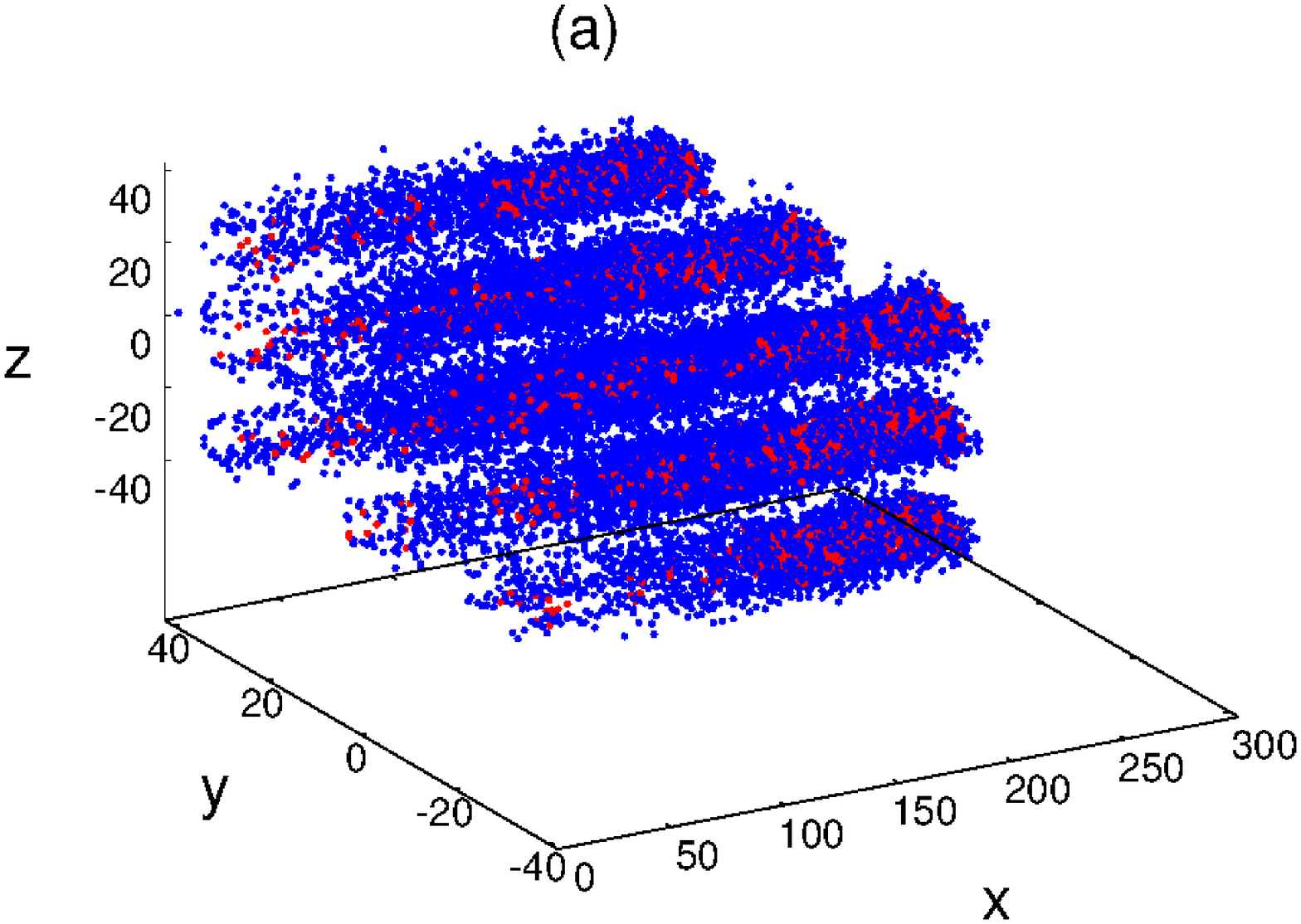}
\includegraphics[width=0.53\columnwidth]{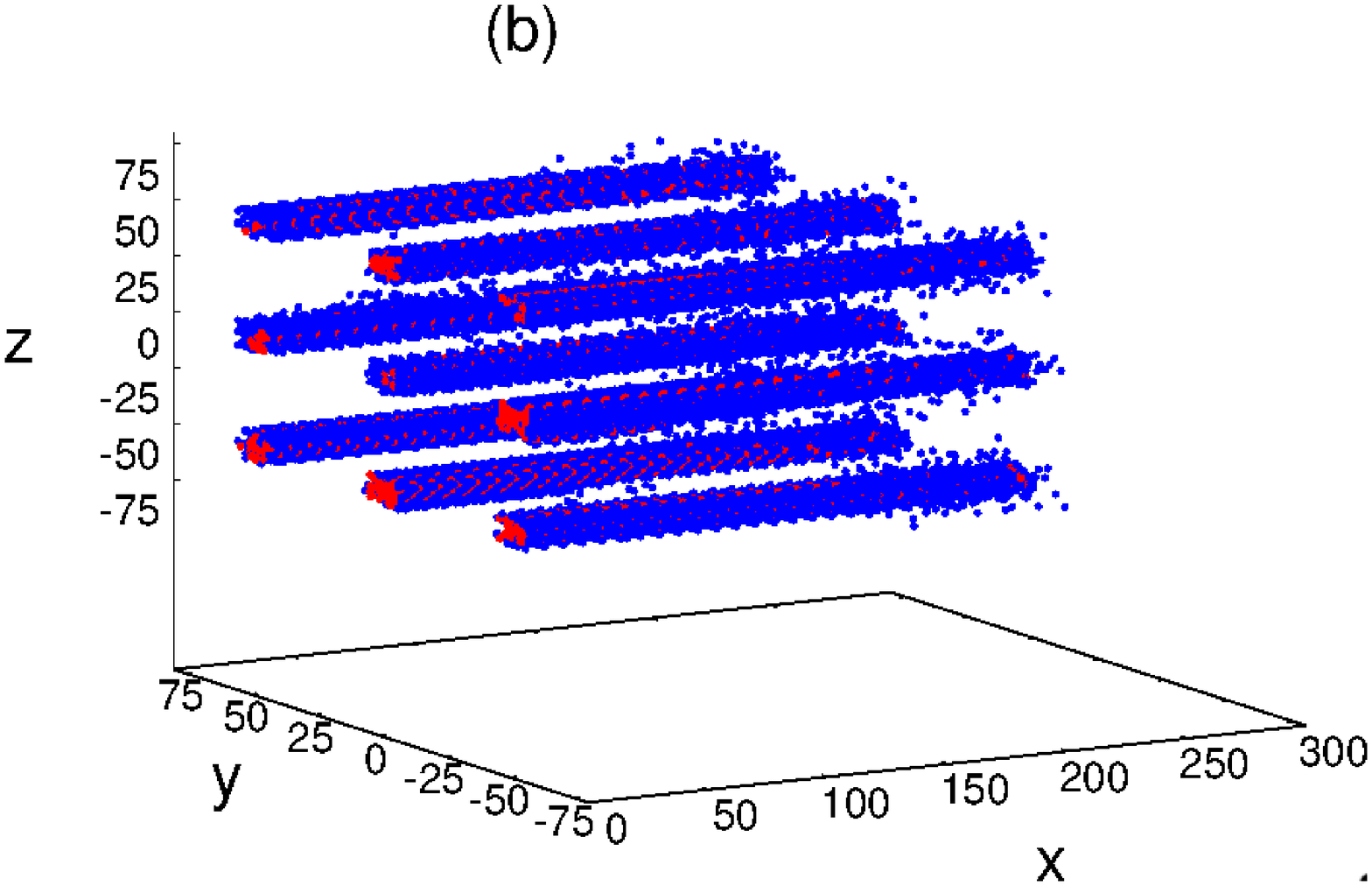}
\caption{\label{fig0} Snapshots of nanotubes (N-T) formed from the exit points of 9 nanochannels (N-C) arranged in a square lattice 
at $x=0$ for two different growth conditions. Blue particles denote C, whereas A is shown as red-dots. B is not shown to maintain clarity.
Plot (a)  $p_A=0.01$, $p_P=0.025$, $D_A=D_C=0.0125$, $D_B=0.05$,  $\rho_B=0.6$, $S_{cyl}=18.$;
(b) $p_A=0.0075$, $p_P=0.0125$, $D_A=D_C=0.0125$, $D_B=0.2$,  $\rho_B=0.6$, $S_{cyl}=40$. Only $1/4$\& $1/8$-th
of A\&C particles are seen.  
}
\end{figure}

\begin{figure}
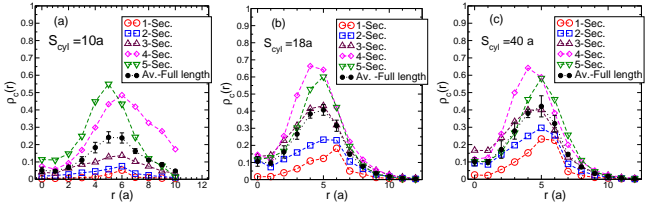

\includegraphics[width=0.32\columnwidth]{S_cyl_10.eps}
\includegraphics[width=0.32\columnwidth]{S_cyl_18.eps}
\includegraphics[width=0.32\columnwidth]{S_cyl_40.eps}
\caption{\label{fig1} Plots of the average number density $\rho_C(r)$ (units of $a^{-3}$)  of C (CdS) of different 
sections along the length of N-T as a function of the 
radial distance $r$  from the center of the N-T. Plots (a), (b) and (c) correspond to 
different separations $S_{cyl}$ between the N-C exits.  
The first section is closest to the N-C exit at $x=0$,
whereas the 5th-section is farthest from the N-C exit and contains the closed end. 
The  density  averaged over all the 5 sections of the N-T is also shown. 
Values of other parameters are $p_A=0.01$, $p_P=0.025$, $D_A=D_C=0.0125$, $D_B=0.05$,  $\rho_B=0.6$.
All parameters are probabilities, thereby have no units. 
}
\end{figure}

Fig. \ref{fig0} shows two snapshots of C-particle N-T from our simulations. The radius of the N-C exits 
are $r_c = 5a$ in both figures, however, the distance $S_{cyl}$ between the N-C walls are correspondingly 
$L_y \times L_z$  are different in Fig. \ref{fig0} $a$ and $b$.  For Fig. \ref{fig0}$a$ $S_{cyl} =18a$,
the distance $d_c$  between N-C centers is $d_c=28a$ and $L_x \times L_y \times L_z = 1000 \times 84 \times 84 a^3$,
whereas for Fig.  \ref{fig0}b $S_{cyl} =40a$,  $d_c=50a$ and $L_x \times L_y \times L_z = 1000 \times 150 \times 150 a^3$
Figure \ref{fig0}(a) shows that the N-Ts are well-formed at one end near the $x=200a$, but the wall has low C-density
near $x=0$, i.e., near the exit of the nano-channel. On the other hand, Fig. \ref{fig0}b, with a higher value
of $D_B$ and $S_{cyl}$ shows clear and well-formed N-Ts with high density of C along the entire length 
of the N-T.
The reasons for such observations have been  discussed earlier: depletion of B-ions supplies near the 
N-C exits as the reaction and growth of N-T gradually proceeds. A higher value of separation $S_{cyl}$ 
between N-T walls and a higher $D_B = 0.2$ naturally ensures a better supply of B to the reactant A-ions near the 
N-C exits leading to better N-Ts, as seen in Fig. \ref{fig0}b.

To separate out  the effects of varying $S_{cyl} $ and $D_B$ in our model we fix a relatively low 
value of the ratio $D_B/D_A=4$ and vary $S_{cyl}$ in subfigures (a), (b) and (c) and plot the average density $\rho_C(r)$
of $C$ as a function of the radial distance $r$ from the center of  the respective N-Ts.
Furthermore, each of the 9 N-Ts are divided into 5 sections along its length, and we calculate and plot the density profile 
in each section (averaged over the 9 N-Ts) as a function $r$.  The quantity $\rho_C(r)$ has low values near N-T center ($r=0$)
but peaks near $r=5a$ indicative of the formation of walls of distinct N-T of C with a pore at center, 
and then decays back nearly to $0$ with increase of $r$. 
A non-zero value of $\rho_C(r)$ near $r \approx 10a$ is indicative of spread of C and the fusion between adjacent N-T,
as seen especially for section 4 with $S_{cyl} =10a$.  The density profile of section-5 (section containing the closed end and farthest 
from the NC-exit) is nearly 
the same for $S_{cyl} =10a, 18a$ and $40a$, but the peak in $\rho_C(r)$ for section-1 gets progressively higher for larger $S_{cyl}$ values.
 We see a higher density of C in section-4 near $r=5a$ in Fig. \ref{fig1} (b) and (c) because just after the initial phase 
of flow of A and plug formation, there will be cylinder of surplus A between plug and N-C exit, which will promptly react 
with B at the surface and aggregate to form N-T walls. 
This will be followed by B-deficiency leading to lower values of $\rho_C(r)$-peak  for sections $3,2$ and $1$.

\begin{figure}
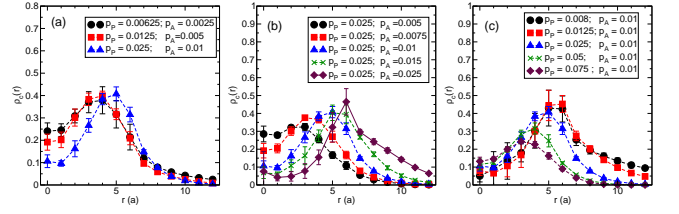

\includegraphics[width=0.32\columnwidth]{chG_ratio_A_G_constant.eps}
\includegraphics[width=0.32\columnwidth]{ch_denA_g025.eps}
\includegraphics[width=0.32\columnwidth]{chG_denA_01.eps}
\caption{\label{fig2} Plots of $\rho_C (r)$ versus $r$. These figures explore the effect of varying the growth rate 
$g = p_P a/\tau$ and the rate of influx of A particles (by varying the probability $p_A$) on the N-T cylinder formation.
In subfigure (a) the ratio $p_P/p_A =2.5$ is held fixed, in (b) $p_P$ is kept fixed while $p_A$ is varied, in (c)
$p_p$ take different values maintaining $p_A = 0.01$. Other parameter values are 
$D_A=D_C=0.0125$, $D_B=0.05$, $\rho_B=0.6$, $S_{cyl}=18a$. The reference case of $p_P =0.025$ and $p_A=0.01$ plotted as blue filled triangles in (b) and (c). 
}
\end{figure}

Since the N-T formation by reaction, diffusion and aggregation is critically dependent on the rate of 
replenishment of A as well as
the growth-rate $g = p_P a/\tau$ of N-T as compared to the diffusion-rate of B. Hence in Fig. \ref{fig2} we keep $D_B$ fixed
and vary the probabilities $p_A$ and $p_P$ to investigate how the density profile $\rho_C(r)$ is affected by the variation. 
Low values of $p_A$ ($p_A = 0.0025, p_A=0.005$) keeping the ratio 
$p_P/p_A = 2.5$ fixed fills up the axial of the N-T at $r=0$ with C, as B get more time to diffuse in and react with A.
For such cases we obtain solid cylinders as seen in Fig. \ref{fig2}a and b. 
In Fig. \ref{fig2}b, lowering the rate of replenishment of A by decreasing 
$p_A$ leads to the shrinking of the radius of N-T seen by the shift in the peak of $\rho_C(r)$, 
whereas, a higher value of $p_A =0.025$ ensures an excess of A at the N-C exit which fills up the center of  the 
N-T with A, and so C is excluded from the axial region. Thus a wide tube with a  pore at the center is obtained. 
Moreover, non-zero values of $\rho_C$ at $r=10a$ indicates that this excess 
A spreads out radially, reacts with B to form C, and gets conjoined with the neighbouring N-T. On the other hand 
in Fig. \ref{fig2}c  with fixed $p_A =0.01$, a low growth rate set by $p_P = 0.008$ again leads to diffussive 
spread of excess A exiting the AAO N-C to form C after reacting with B-s at radii $r>6 a$ , thus again leading to the possibility of fused 
N-T for $S_{cyl} = 18 a$.    Higher growth rate of N-T lead to the shift in the peak of $\rho_C (r)$ radially inwards.
Thus a balance of the values of $p_P $ and independently $p_A$  are essential for the growth of well-separated distinct
N-Ts. Of course experimentally, $p_P$ and $p_A$ are not free parameters as  in simulations, but some control and 
variation over these quantities could be achieved experimentally by playing around with the material of the N-C (in this case AAO)
and/or the radius of the N-C.

\begin{figure}
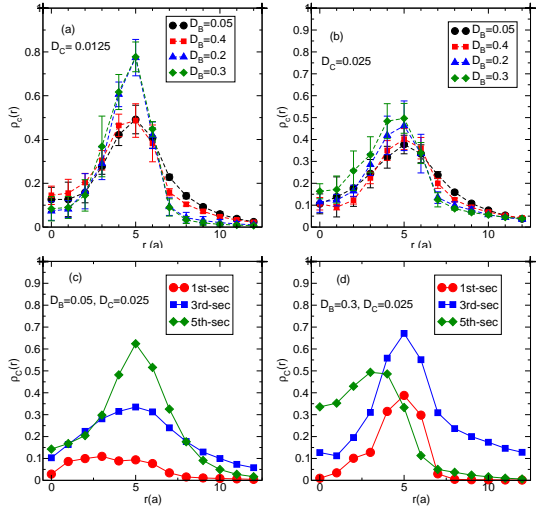

\includegraphics[width=0.4\columnwidth]{ch_diffB_DC0125.eps} 
\includegraphics[width=0.4\columnwidth]{ch_diffB_DC025.eps} \\
\includegraphics[width=0.4\columnwidth]{sec_diffB_05_DC025.eps} 
\includegraphics[width=0.4\columnwidth]{sec_diffB_3_DC025.eps} \\
\caption{\label{fig3} Plots of $\rho_C (r)$ versus $r$. Subfigure (a) and (b) explore the role 
of varying rate of diffusion of B ions (controlled by probability $D_B$) on cylinder formation at two different values of 
the probability $D_C$  of diffusive random hop of C-particle.  Subfigures (c) and (d) show the  average density of different sections
 along the length of the N-T at two different values of $D_B$. Other values are $p_A=0.0075$, $p_P=0.0125$, $D_A=0.0125$, $\rho_B=0.6$, $S_{cyl}=18a$.
}
\end{figure}

Next, in Fig. \ref{fig3}a and \ref{fig3}b we observe the change in density profile $\rho_C (r)$ with varying 
$D_B$ for two different values of $D_C$. Higher values of $D_B$ ($D_B = 0.2,0.3$) lead to sharp peaks  
in the density for $C$ at $r\approx 5a$ pointing to well separated distinct N-T, refer  Fig. \ref{fig3}a. 
However, increasing $D_C$ leads to diffusive spread of C before setting down to final position within N-T
leading to a lower value of maxima of $\rho_C(r)$, refer Fig. \ref{fig3}b. But even in this case a large 
assymetry in the values of $D_B$ and $D_A$ favours high density of C at the N-T walls. There is a jump in 
the maximum of $\rho_C(r)$ from $0.5 a^{-3}$ to $0.8 a^{-3}$ as $D_B$ is changed from $0.1$ to $0.2$ corresponding to 
$D_B/D_A$ change from $8$ to $16$, but does not change significantly thereafter. We has shown 
earlier \cite{kiruthiga} that $D_C=0$ is not a amenable condition for good growth of N-T walls, thus 
a small but finite value of $D_C$ plays a crucial role in the self-assembly.

Figures \ref{fig3}c and \ref{fig3}d  focus at the $\rho_C(r)$ for different sections 
along the length of the N-Ts  for two distinct values
of $D_B$ at $D_C=0.025$ . The 5th section always has a non-zero density of C  at $r=0$, refer both 
\ref{fig3}c and \ref{fig3}d. The 3-rd section at the middle of the N-T has a higher density of C than 
5th section in Fig. \ref{fig3}d for $D_B =0.3$. Even the 1-st section in Fig. \ref{fig3}d
manages a distinct peak corresponding to a dense wall of C particles at $r=5a$.  In contrast,  we see in  Fig. \ref{fig3}c
that the 3rd section already feels B-deficiency near the N-C exit for  $D_B =0.05$, with lower values of 
$\rho_C(r)$ in section-3 than section-5. The values of $\rho_C(r)$ are close to zero in the 1-st section near N-C exit,
indicating an acute scarcity of B-ions. We see also a significant spread in the density profile of 
$C$ at sections $5$ and $3$, thus forming fused N-Ts.

\begin{figure}
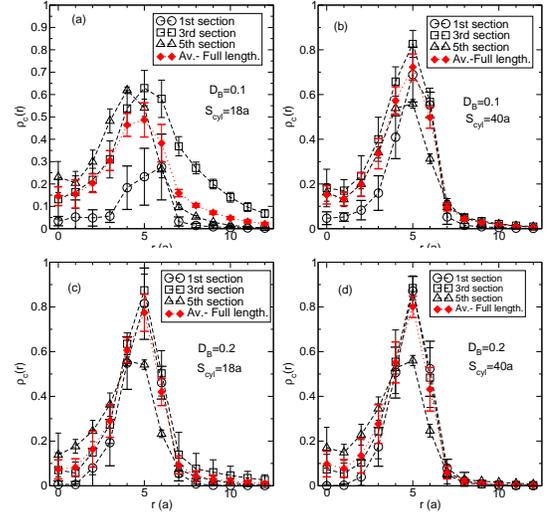

\includegraphics[width=0.4\columnwidth]{diffB_1.eps}
\includegraphics[width=0.4\columnwidth]{diffB_1_large_box.eps} \\
\includegraphics[width=0.4\columnwidth]{diffB_2.eps} 
\includegraphics[width=0.4\columnwidth]{diffB_2_large_box.eps}
\caption{\label{fig4}  Plots of $\rho_C(r)$ versus $r$. These subfigures explore the av. density 
variation at different sections along the length of tube for two different values of separation $S_{cyl}$  
between N-T and two different values of $D_B$. In contrast to Fig. \ref{fig1}, $D_B/D_A \approx 10$ in this figure.
Other parameters are kept fixed at $p_A=0.0075$, $p_P=0.0125$, $D_A=D_C=0.0125$, $\rho_B=0.6$.
}
\end{figure}

The plots of Fig. \ref{fig4} further analyze the data of Fig.\ref{fig3}a, but 
with added perspective of having two different values of $S_{cyl}$ for $D_B=0.1$ and $D_B =0.2$.
For $S_{cyl}=18a$ and $D_B=0.1$ in Fig. \ref{fig4}a, the 5th and 3rd section shows a peak around $0.6 a^{-3}$,
but $\rho_C(r)$ for section-2 has a wider spread than for section-5, indicating that A has  to diffuse further out
before it meets B to form C. Section-1 of Fig. \ref{fig4}a shows low values of $\rho_C(r)$,  
but for $S_{cyl}=40a$ in  Fig. \ref{fig4}b $\rho_C(r)$ for all sections show a significant jump in value.
Moreover, the ratio of the maximum-values of  $\rho_C$ for  section-1 and 5 is lower for  Fig. \ref{fig4}b,
indicating relatively uniform densitites along the length of the N-Ts. The growth of N-Ts of 
uniform densities is further improved in Fig. \ref{fig4}c and d for $D_B =0.2$ with $\rho_C \approx 1 a^{-3}$ 
at the walls.  The values of the maxima of  $\rho_C(r)$ is nearly the same in both Fig. \ref{fig4}c and d. 
This is in contrast with Fig. \ref{fig4}a and b,
where the peak in the average $\rho_C(r)$ (over all sections) is higher in Fig. \ref{fig4}b compared to a.   
Higher values of B-diffusivities ensure a good supply of B from bulk, thereby $\rho_C(r)$ in Fig. \ref{fig4}c and d
do not crucially depend on $S_{cyl}$-values. Of course, distinct N-Ts of (nearly) uniform densities do not form for $S_{cyl}=10 a$.

In conclusion, a improved crop of self-assembling $CdS$ (C) nano-tubes can be obtained by 
increasing the distance $S_{cyl}$ between the N-Cs in the AAO template. This allows
larger amounts  of $ {\rm S^{-2}}$ (B) to diffuse in between the growing CdS N-Ts.
Alternatively, a lower rate of inflow of A (${\rm Cd^{+2}}$)  ions into the B (${\rm S^{2-}}$)
chamber promotes uniform growth, but have to be suitably balanced by growth rate $p_P a/\tau$,
else one obtains solid cylinders without a axial pore, or else, fused nano-tubes.
Furthermore, we predict that good growth of cylindrical C nano-tubes with high-density
walls will be occur  only when there is marked assymmetry in the  diffusion constants of $A$ 
and $B$ ions with $D_B \gg D_A$ and  a non-zero, small value of $D_C$. Experimentally,
an excess of Cd on the surface of CdS N-Ts have been observed in tune with our 
expectations \cite{shouvik1},
but stoichiometric analysis of CdS densities at different sections of the N-Ts are yet to be done.

AC thanks computational facilities of  Nano-Science unit at IISER,  funded 
by DST, India: project no. SR/NM/NS-42/2009, and discussions with Shouvik Datta. 
JK acknowledges the summer research fellowship to visit IISER-Pune provided by IAS, Bangalore, India.

\newpage



\begin{thebibliography}{99}


\bibitem{kiruthiga} J. Kiruthiga and A. Chatterji, J. Chem. Phys., {\bf 138} 024905 (2013).
\bibitem{shouvik}
A. Varghese, S. Datta, Phys. Rev. E., {\bf 85}, 056104 (2012).
\bibitem{israel} Intermolecular and Surface Forces, J.N. Israelachvili, 3rd Edition, Academic Press, (2010).
\bibitem{vermant} M. Grzelczak, J. Vermant, E.M.Furst, L.M.Liz-Marzan, ACS-Nano,{\bf 4},3591 (2010).
\bibitem{witten} T.A. Witten, Rev. Mod. Phys., {\bf 71}, S368 (1999).
\bibitem{biop} A.J. Koch and H. Meinhardt, Rev. Mod. Phys., {\bf 66}, 1481 (1994).
\bibitem{einax} M. Einax, W. Dieterich, P. Maass,  Rev. Mod. Phys., {\bf 85} 921 (2013). 
\bibitem{opto} K. Dholakia and P. Zemanek,  Rev. Mod. Phys., {\bf 82} 1767 (2013).
\bibitem{majumdar} M. Majumdar, N. Chopra, R. Andrews and B.J. Hinds, Nature, {\bf 438}, 44, (2005). 
\bibitem{shouvik1} Private communications with Shouvik Dutta.
%
\end{thebibliography}
\end{document}